\documentclass[pra,twocolumn,superscriptaddress,showpacs,nofootinbibfloatfix,amsmath,amsfonts,amssymb]{revtex4-1}%
\usepackage{amsmath,amsfonts,amssymb,color}
\usepackage{amsthm}
\usepackage{leftidx}
\usepackage{graphicx}
\usepackage{xcolor}
\usepackage{dcolumn}
\usepackage{bm}
\usepackage{epstopdf}
\usepackage{epsfig}
\usepackage{environ}
\usepackage{pdfcomment}

\usepackage{float}
\usepackage[T1]{fontenc}
\usepackage[latin9]{inputenc}
\usepackage{setspace}
\usepackage{esint}

\begin{document}

\preprint{APS/123-QED}

\title{Non-Hermitian Floquet topological phases in the double-kicked rotor}

\author{Longwen Zhou}
\email{zhoulw13@u.nus.edu}
\affiliation{%
	Department of Physics, College of Information Science and Engineering, Ocean University of China, Qingdao, China 266100
}
%\affiliation{%
%	Department of Physics, National University of Singapore, Singapore 117543
%}
\author{{Jiaxin Pan}}
\affiliation{%
	Department of Physics, College of Information Science and Engineering, Ocean University of China, Qingdao, China 266100
}
%\author{Jiangbin Gong}%
%\email{phygj@nus.edu.sg}
%\affiliation{%
%	Department of Physics, National University of Singapore, Singapore 117543
%}

\date{\today}

\begin{abstract}
Dynamical kicking systems possess rich topological structures. In this work,
we study Floquet states of matter in a non-Hermitian extension of double kicked
rotor model. Under the on-resonance condition, we find various non-Hermitian
Floquet topological phases, with each being characterized by a pair of topological
winding numbers. A generalized mean chiral displacement is introduced to detect these winding numbers dynamically in two symmetric time frames.
Furthermore, by mapping the system to a
periodically quenched lattice model, we obtain the topological edge
states and unravel the bulk-edge correspondence of the non-Hermitian double kicked rotor. These
results uncover the richness of Floquet topological
states in non-Hermitian dynamical kicking systems.
\end{abstract}

%Use showkeys class option if keyword
\maketitle

\section{Introduction}\label{sec:Int}

Floquet topological phases of matter emerge in systems under time-periodic modulations. 
One class of Floquet systems that has been shown to possess rich topological properties are dynamical kicking systems~\cite{LeboeufPRL1990}. They are first introduced in the study of dynamical localization and quantum chaos, with the kicked rotor (KR) been a prototypical example~\cite{KR1,KR2,KR3,KR5,KR6,KRRevs}. In 2008, Wang and Gong analyzed a modified version of the KR~(also called double kicked rotor)~\cite{DKRGong}, and discovered its fractal
quasienergy spectrum that mimicking the Hofstadter butterfly in quantum
Hall effects~\cite{Butterfly}. Later, rich Floquet topological states in the double
kicked rotor (DKR) were characterized, and then employed to achieve
quantized acceleration in momentum space~\cite{DerekORDKR}. The topological equivalence
between the DKR and the kicked Harper model~\cite{PLADana}, another prototypical
dynamical kicking system, has also been proved rigorously~\cite{HailongPRE2013}. The introduction of a spin-$1/2$ degree of freedom to KR and DKR further uncovers the richness of Floquet topological states that can appear in dynamical kicking systems~\cite{ZhouPRA2018,KR4,DahlhausPRB2011}.

In the past decade, Floquet topological phases have attracted great interest across a broad range of research areas. This is mainly due to the richness and high-tunability of their topological properties~\cite{OkaPRB2009,LindnerNP2011,KitagawaPRB2011,LeonPRL2013,CayssolRRL2013,ThakurathiRRB2013,GrushinPRL2014,Wang2014,TitumPRL2015,DLossPRLPRB,KitagawaPRB2010,JiangPRL2011,KunduPRL2013,Bomantara2016,ZhaoErH2014,ReichlPRA2014,LeonPRB2014,AnomalousESPRX,FulgaPRB2016,ZhouPRB2016,AsbothSTF,TongPRB2013,ZhouEPJB2014,XiongPRB2016,SeradjehArXiv2017,YapPRB2017,YapPRB2018,ZhouPRB2018,LinhuarXiv2018}, with potential applications in ultrafast electronics~\cite{OkaRev}, quantum simulation~\cite{EckardtRMP2017} and quantum computing~\cite{BomantaraFloquetQC2018}. The topological classification of these dynamical states of matter also require new schemes~\cite{NathanNJP2015,ClassificationFTP1,ClassificationFTP2} that go beyond their static cousins. Experimentally, Floquet topological phases have been realized in cold atom, photonic, phononic and acoustic systems~\cite{ColdAtomFTP,ColdAtomFTP2,PhotonFTP0,PhotonFTP1,PhotonFTP2,PhononFTP,PhononFTP2}.

In recent years, the study of Floquet topological phases has been
extended to non-Hermitian domain~\cite{NHTPReview,NHTPReview2}.
There, gain and loss or nonreciprocal effects were introduced to make the evolution of Floquet systems nonunitary~\cite{YuceFloquetPT,KimArXiv2016,GongPRA2015,WuPRL2015,WuPRA2019}.
In quantum walk setups, gain and loss were implemented in several
studies to measure the topological invariants~\cite{RudnerPRL2009,ZeunerPRL2015,MochizukiPRA2016,HuangPRA2016,RakovszkyPRB2017,XiaoNatPhys2017,ChenPRA2018,HarterArXiv2018,WangArXiv2018}. Furthermore, a periodically
quenched nonreciprocal lattice model has been found to possess abundant
Floquet topological phases with arbitrarily many topological edge
states induced by non-Hermitian effects~\cite{ZhouNHDQL2018}. In dynamical kicking systems,
a ${\cal PT}$-symmetric kicked rotor was proposed~\cite{WestPTKR2010,PTKR2017} and its transport
properties have been investigated in \cite{ZhaoPTKR2019}. However, the
richness of non-Hermitian Floquet topological phases in dynamical
kicking systems have not been revealed yet.

In this work, we introduce a DKR with complex kicking strengths,
and unravel its fruitful non-Hermitian Floquet topological phases.
After introducing our model in Sec.~\ref{sec:Model}, we analyze its spectrum, symmetry
and topological properties in Sec.~\ref{sec:NH-ORDKR}. A pair of integer winding numbers is introduced to fully characterize the topological phases appearing in the non-Hermitian DKR. We further extend the definition of mean chiral displacement (MCD) to nonunitary evolution,
and using it as a probe to extract the topological winding numbers
of non-Hermitian DKR dynamically. By mapping our system to a periodically
kicked lattice model, we also present its topological edge states
under open boundary condition (OBC) and demonstrate its bulk-edge correspondence.
We conclude our work and discuss potential future directions in Sec.~\ref{summary}. 

\section{The model}\label{sec:Model}
The DKR model is described by the Hamiltonian ${\hat H}=\frac{{\hat p}^2}{2}+\kappa_1\cos({\hat x}+\beta)\sum_{\ell\in{\mathbb Z}}\delta(t-\ell T)+\kappa_2\cos({\hat x})\sum_{\ell\in{\mathbb Z}}\delta(t-\ell T-\tau)$. It can be realized by cold atoms subject to counter-propagating laser pulses in an optical lattice~\cite{DKREXP,DKREXP2,DKREXP3}, where $\hat{x}$ and $\hat{p}$ are position and momentum operators of cold atoms. In a driving period $T$, the system is first kicked by a lattice potential of strength $\kappa_{1}$. Then it is evolved freely over a time duration $\tau\in(0,T)$, kicked by another lattice potential of strength $\kappa_{2}$, and then evolved freely over another time duration $T-\tau$. $\beta$ is a controllable phase shift between the two kicking potentials. The Floquet operator of DKR, obtained by integrating the Schr\"odinger equation $i\hbar\partial_t|\psi\rangle={\hat H}|\psi\rangle$ over a complete driving period~(e.g., from $t=\ell T-0^+$ to $t=(\ell+1)T-0^+$), is given by
\begin{equation}
	\hat{U}=e^{-i(T-\tau)\frac{\hat{p}^{2}}{2\hbar}}e^{-i\frac{\kappa_{2}}{\hbar}\cos(\hat{x})}e^{-i\tau\frac{\hat{p}^{2}}{2\hbar}}e^{-i\frac{\kappa_{1}}{\hbar}\cos(\hat{x}+\beta)}.\label{eq:DKRM}
\end{equation}
In the Floquet operator, the spatial periodicity of kicking potentials
allow the momentum $\hat{p}$ to take eigenvalues $p=(n+\eta)\hbar$,
where $n\in\mathbb{Z}$ and $\eta\in(0,1)$ being the conserved quasimomentum.
For a Bose-Einstein condensate of large coherence width, one can choose $\eta=0$~\cite{OnRes1,BetaEqT0}. The
momentum $\hat{p}$ is then quantized as $\hat{p}=\hat{n}\hbar$,
i.e., integer multiples of the effective Planck constant $\hbar$.
Furthermore, under the condition $\hbar T=4\pi$~\cite{OnRes1,BetaEqT0,OnRes2}, we obtain the on-resonance
DKR (ORDKR) model, whose Floquet operator takes the form
\begin{equation}
	\hat{U}=e^{+i\frac{\hbar\tau}{2}\hat{n}^{2}}e^{-iK_{2}\cos(\hat{x})}e^{-i\frac{\hbar\tau}{2}\hat{n}^{2}}e^{-iK_{1}\cos(\hat{x}+\beta)}.\label{eq:ORDKR}
\end{equation}
Here $K_{1}=\kappa_{1}/\hbar$ and $K_{2}=\kappa_{2}/\hbar$ represent dimensionless kicking strengths. It has
been shown that this ORDKR model possesses rich topological properties,
including the Hofstadter butterfly-like Floquet spectrum~\cite{DKRGong}, quasienergy
bands with large Chern numbers, and quantized Thouless
pumping in momentum space~\cite{DerekORDKR}.

In this work, we further investigate the Floquet topological phases
of ORDKR in non-Hermitian regime. More specifically, we will focus
on the two-band situation by choosing the time delay $\tau$ between
the two kicks such that $\hbar\tau=\pi$. The resulting non-Hermitian
ORDKR (NH-ORDKR) model is described by the Floquet operator 
\begin{equation}
	\hat{U}=e^{i\frac{\pi}{2}\hat{n}^{2}}e^{-iK_{2}\cos(\hat{x})}e^{-i\frac{\pi}{2}\hat{n}^{2}}e^{-iK_{1}\cos(\hat{x}+\beta)},\label{eq:NH-ORDKR}
\end{equation}
where the kicking strengths
\begin{equation}
	K_{j}=u_{j}+iv_{j}\qquad j=1,2\label{eq:K12}
\end{equation}
now take complex values, with $\{u_{1},v_{1},u_{2},v_{2}\}\in\mathbb{R}$.
For an optical lattice, the imaginary parts of kicking strengths correspond to
particle losses, which may be generated by using a resonant optical beam to kick the
atoms out of the trap. It may also be realized by applying a radio frequency pulse to excite atoms to an irrelevant state, leading to an effective decay when atoms in that state experience a loss by applying an antitrap~\cite{XuPRL2017}.
In photonic systems, a complex kicking strength correspond to a
complex refractive index, whose imaginary part represents either loss
or gain. 
This kind of potential has interesting engineering applications, such as realizing unidirectional transport of light~\cite{LonghiPRA2010} and other types of laser devices~\cite{WuPRL2015}.
In the following,
we will unravel rich Floquet topological phases in the NH-ORDKR induced
by complex kicking lattice potentials.

\section{Floquet topological phases in NH-ORDKR}\label{sec:NH-ORDKR}

In this section, we first analysis the Floquet operator of NH-ORDKR in Eq.~(\ref{eq:NH-ORDKR}), and discuss the symmetry that protects
its topological properties. Next, we investigate the quasienergy spectrum
and the conditions of topological phase transitions in the NH-ORDKR.
A pair of integer topological winding numbers is introduced to
characterize each of its Floquet topological phases. To detect these
winding numbers and distinguish different Floquet topological phases
in the NH-ORDKR experimentally, we suggest to measure the MCD
of a wave packet in the optical lattice. Finally, we
map the Floquet operator of NH-ORDKR to a kicked lattice model in
position representation, and uncover its Floquet edge states and
bulk-edge correspondence under OBC.

\subsection{Floquet operator and chiral symmetry}\label{sec:UandCS}

The Floquet operator of NH-ORDKR, as defined in Eq.~(\ref{eq:NH-ORDKR}),
is translational invariant over two sites (i.e., $\hat{n}\rightarrow\hat{n}+2$) in the momentum lattice. By introducing
a bipartite lattice basis in momentum space and taking the
periodic boundary condition, we could express the Floquet operator
of NH-ORDKR as $\hat{U}=\sum_{\theta}U(\theta)|\theta\rangle\langle\theta|$,
where 
\begin{alignat}{1}
	U(\theta)= & e^{+i\frac{\pi}{4}\sigma_{z}}e^{-i{\cal K}_{2}\left(\cos\frac{\theta}{2}\sigma_{x}+\sin\frac{\theta}{2}\sigma_{y}\right)}\nonumber \\
	\times & e^{-i\frac{\pi}{4}\sigma_{z}}e^{+i{\cal K}_{1}\left(\cos\frac{\theta}{2}\sigma_{x}+\sin\frac{\theta}{2}\sigma_{y}\right)},\label{eq:UTheta}
\end{alignat}
and
\begin{equation}
	{\cal K}_{1}\equiv K_{1}\sin\frac{\theta}{2},\qquad{\cal K}_{2}\equiv K_{2}\cos\frac{\theta}{2},\label{eq:K1K2}
\end{equation}
with $\theta\in[-\pi,\pi)$ being the conserved quasiposition due
to translational symmetry in momentum space, and $\sigma_{x,y,z}$
being Pauli matrices in their usual representation {[}see Appendix
\ref{sec:App-A} for derivation details of Eq. (\ref{eq:UTheta}){]}.
We have also set the phase delay between two kicks to be $\beta=\frac{\pi}{2}$,
which allows $U(\theta)$ to possess nontrivial topological phases
when $K_{1,2}$ taking real values~\cite{DerekORDKR}. 

To characterize the symmetry and topological properties of $U(\theta)$,
we introduce a pair of symmetric time frames by resetting the start
time of the evolution. In these time frames, $U(\theta)$ takes the
form
\begin{alignat}{1}
	U_{1}(\theta)= & e^{-i\frac{{\cal K}_{2}}{2}\left(\cos\frac{\theta}{2}\sigma_{x}+\sin\frac{\theta}{2}\sigma_{y}\right)}e^{-i{\cal K}_{1}\left(\sin\frac{\theta}{2}\sigma_{x}-\cos\frac{\theta}{2}\sigma_{y}\right)}\nonumber \\
	\times & e^{-i\frac{{\cal K}_{2}}{2}\left(\cos\frac{\theta}{2}\sigma_{x}+\sin\frac{\theta}{2}\sigma_{y}\right)},\label{eq:U1Theta}
\end{alignat}
\begin{alignat}{1}
	U_{2}(\theta)= & e^{+i\frac{{\cal K}_{1}}{2}\left(\cos\frac{\theta}{2}\sigma_{x}+\sin\frac{\theta}{2}\sigma_{y}\right)}e^{-i{\cal K}_{2}\left(\sin\frac{\theta}{2}\sigma_{x}-\cos\frac{\theta}{2}\sigma_{y}\right)}\nonumber \\
	\times & e^{+i\frac{{\cal K}_{1}}{2}\left(\cos\frac{\theta}{2}\sigma_{x}+\sin\frac{\theta}{2}\sigma_{y}\right)}.\label{eq:U2Theta}
\end{alignat}
Note that both $U_{1}(\theta)$ and $U_{2}(\theta)$ are similar to
$U(\theta)$ (see Appendix \ref{sec:App-A} for more details). Therefore,
they share the same Floquet spectrum with $U(\theta)$
even if $K_{1}$ and $K_{2}$ are complex numbers. Furthermore, under
the unitary transformation $\Gamma=\sigma_{z}$, we have
\begin{equation}
	\Gamma U_{\alpha}(\theta)\Gamma=U_{\alpha}^{-1}(\theta)\qquad\alpha=1,2,\label{eq:U12CS}
\end{equation}
which means that $U_{1}(\theta)$ and $U_{2}(\theta)$ have the chiral
(sublattice) symmetry $\Gamma$. According to the symmetry classification
of chiral symmetric Floquet systems in one-dimension~\cite{AsbothSTF} and its extension
to non-Hermitian systems~\cite{ZhouNHDQL2018}, each topological phase of $U(\theta)$ can
be described by a pair of integer winding numbers extracted from
$U_{1}(\theta)$ and $U_{2}(\theta)$. We will analyze the spectrum
and topological properties of the NH-ORDKR in detail in the following
subsections.

\subsection{Quasienergy dispersion, topological invariants and phase diagram}\label{sec:PhsDiag}

Expanding $U_{1}(\theta)$ and $U_{2}(\theta)$ by the Euler formula,
and recombining the resulting terms, we can express Eqs. (\ref{eq:U1Theta})
and (\ref{eq:U2Theta}) in a compact form as
\begin{equation}
	U_{\alpha}(\theta)=e^{-iE(\theta)(n_{\alpha x}\sigma_{x}+n_{\alpha y}\sigma_{y})},\label{eq:UAlphaTheta}
\end{equation}
where $\alpha=1,2$ and 
\begin{equation}
	E(\theta)=\arccos(\cos{\cal K}_{1}\cos{\cal K}_{2})\label{eq:Quasienergy}
\end{equation}
gives the quasienergy dispersion relation $\pm E(\theta)$.
Since the real part of $E(\theta)$ is only defined modulus $2\pi$, the quasienergy
spectrum gap closes when ${\rm Im}E(\theta)=0$ and ${\rm Re}E(\theta)=0$ or $\pm\pi$.
When the spectrum becomes gapless, a topological phase transition
may happen. Furthermore, $(n_{\alpha x},n_{\alpha y})$ forms a complex-valued
vector with $n_{\alpha x}^{2}+n_{\alpha y}^{2}=1$ for $\alpha=1,2$
(see Appendix \ref{sec:App-B} for more details). Using these vectors,
we can define a winding number for $U_{\alpha}(\theta)$ as
\begin{equation}
	\nu_{\alpha}=\int_{-\pi}^{\pi}\frac{d\theta}{2\pi}({\bf n}_{\alpha}\times\partial_{\theta}{\bf n}_{\alpha})_z,\quad\alpha=1,2.\label{eq:W12}
\end{equation}
It is not hard to see that $\nu_\alpha$ take real values, as the imaginary part of ${\bf n}_{\alpha}\equiv(n_{\alpha x},n_{\alpha y})$ has no winding in the Brillouin zone~\cite{ZhouNHDQL2018}. Then, following the description of chiral symmetric
non-Hermitian Floquet systems~\cite{ZhouNHDQL2018},
the topological phases of $U(\theta)$ can be characterized by a pair
of integer winding numbers, given by~\cite{Note1}
\begin{equation}
	\nu_{0}=\frac{\nu_{1}+\nu_{2}}{2},\qquad\nu_{\pi}=\frac{\nu_{1}-\nu_{2}}{2}.\label{eq:W0P}
\end{equation}
In the Hermitian limit (imaginary parts of the
two kicking strengths $v_{1}=v_{2}=0$), $\nu_{0}$ and $\nu_{\pi}$ also predict the number of topological edge modes at quasienergy zero and $\pi$
in the ORDKR model~\cite{DerekORDKR}. 

In the following, we will analyze the spectrum and topological phases
of ORDKR under three representative non-Hermitian kicking potentials: 
(i) $v_{1}\neq0,v_{2}=0$ or vice versa; 
(ii) $v_{1}=v_{2}=v\neq0$; and 
(iii) $v_{1}\neq v_{2}$ with $v_{1},v_{2}\neq0$. 
In each case, we give the condition of topological phase transition,
computing winding numbers for each of the topological phases, and
constructing the corresponding topological phase diagram.

\subsubsection{Case (i)}

We first consider the case when only one of the kicking strengths ($K_{1}$ or $K_{2}$)
in Eq. (\ref{eq:K12}) is complex. Under the gapless condition
$\cos[E(\theta)]=\pm1$, it can be
shown that if $v_{1}\neq0$ and $v_{2}=0$, $u_{1},u_{2}$ and $v_{1}$
in Eq. (\ref{eq:K12}) satisfy the equation
(see Appendix \ref{sec:App-C} for derivation details):
\begin{equation}
	v_{1}=\frac{u_{1}}{n\pi}{\rm arccosh}\left[\frac{\pm1}{\cos\left(u_{2}\sqrt{1-\frac{n^{2}\pi^{2}}{u_{1}^{2}}}\right)}\right],\quad n\in\mathbb{Z}.\label{eq:v1Cond}
\end{equation}
Similarly, if $v_{2}\neq0$ and $v_{1}=0$, the gapless condition yields
\begin{equation}
	v_{2}=\frac{u_{2}}{n\pi}{\rm arccosh}\left[\frac{\pm1}{\cos\left(u_{1}\sqrt{1-\frac{n^{2}\pi^{2}}{u_{2}^{2}}}\right)}\right],\quad n\in\mathbb{Z}.\label{eq:v2Cond}
\end{equation}
Note that Eqs.~(\ref{eq:v1Cond}) and (\ref{eq:v2Cond}) are symmetric under the exchange of
subindices $1\leftrightarrow2$. So we can focus on the non-Hermitian
Floquet topological phases and phase transitions related to only one
of them without loss of generality. 

\begin{figure}
	\begin{centering}
		\includegraphics[scale=0.5]{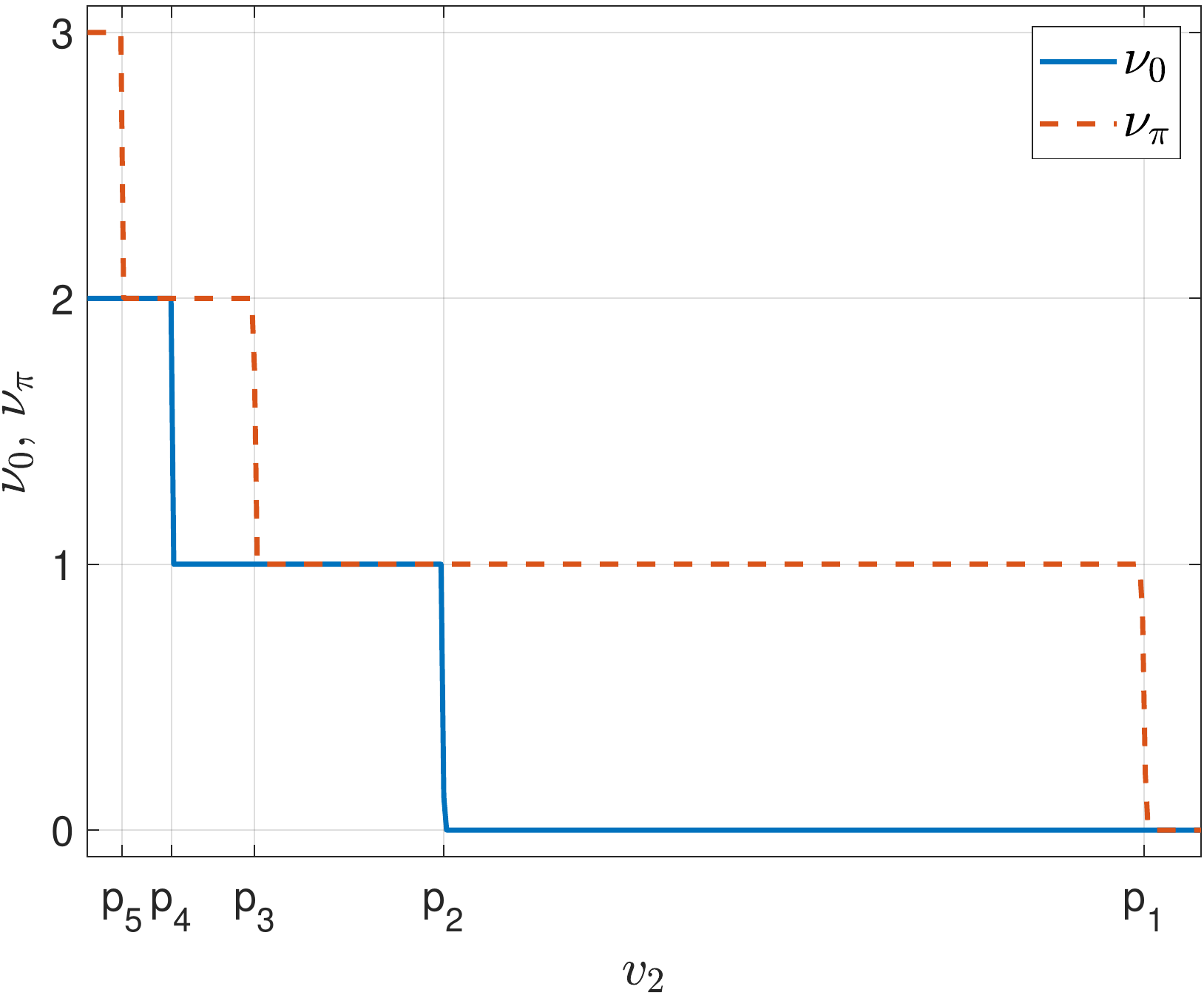}
		\par\end{centering}
	\caption{Evolution of the winding numbers $\nu_{0}$ (blue solid line) and
		$\nu_{\pi}$ (red dashed line) vs. the imaginary part of the kicking
		strength $K_{2}=u_{2}+iv_{2}$. System parameters are chosen as $u_{1}=0.5\pi$
		and $u_{2}=5.5\pi$ and $v_{1}=0$. The numerical values of $p_{1},p_{2},p_{3},p_{4},p_{5}$
		along the $v$-axis are obtained analytically from Eq. (\ref{eq:v2Cond})
		with $n=1,2,3,4,5$.\label{fig:W_vs_v2}}
\end{figure}

To check whether a non-vanishing imaginary part of $K_{1}$ or $K_{2}$
could induce new topological phases in the NH-ORDKR, we need to investigate
the behavior of winding numbers Eq. (\ref{eq:W12}) versus this imaginary
part. A representative example is shown in Fig. \ref{fig:W_vs_v2},
where we choose $u_{1}=0.5\pi$ and $u_{2}=5.5\pi$ for the real parts
of kicking strengths. According to Ref.~\cite{DerekORDKR}, this choice leads
to a Floquet topological phase with $(\nu_{0},\nu_{\pi})=(2,3)$ in
the Hermitian limit. In Fig. \ref{fig:W_vs_v2}, we observe that with
the increase of imaginary kicking strength $v_{2}$, a series of topological phase transitions
happen at $v_{2}=p_{n}$ with $n=1,...,5$ in Eq. (\ref{eq:v2Cond}).
Each transition is accompanied by the vanishing of a spectrum gap, 
together with the quantized change
of winding number $\nu_{0}$ (blue solid line) or $\nu_{\pi}$ (red dashed
line) by $1$. In the limit $v_{2}\rightarrow\infty$, the system
ends in a topologically trivial phase with $\nu_{0}=\nu_{\pi}=0$. 

Therefore, we conclude that a non-vanishing imaginary part in the
kicking strength $K_{1}$ or $K_{2}$ of the NH-ORDKR could indeed
induce topological phase transitions and create non-Hermitian Floquet topological
phases, with each characterized by a pair of integer quantized winding
numbers $(\nu_{0},\nu_{\pi})$. In more general situations, analytical
solutions for the gap closing conditions like Eqs. (\ref{eq:v1Cond})
and (\ref{eq:v2Cond}) may not be available. We will consider these
cases in the following.

\subsubsection{Case (ii)}

In this case, both the two kicking strengths $K_{1}$ and $K_{2}$
take complex values under the constraint that their imaginary parts
are equal, i.e., $v_{1}=v_{2}=v$. Using the gapless condition (see
Appendix \ref{sec:App-C} for more details) and the winding numbers
$(\nu_{0},\nu_{\pi})$, we could then numerically characterize the Floquet
topological phases of NH-ORDKR at different imaginary
kicking strength $v$. Two representative examples will be discussed
as follows.

\begin{figure}
	\begin{centering}
		\includegraphics[scale=0.5]{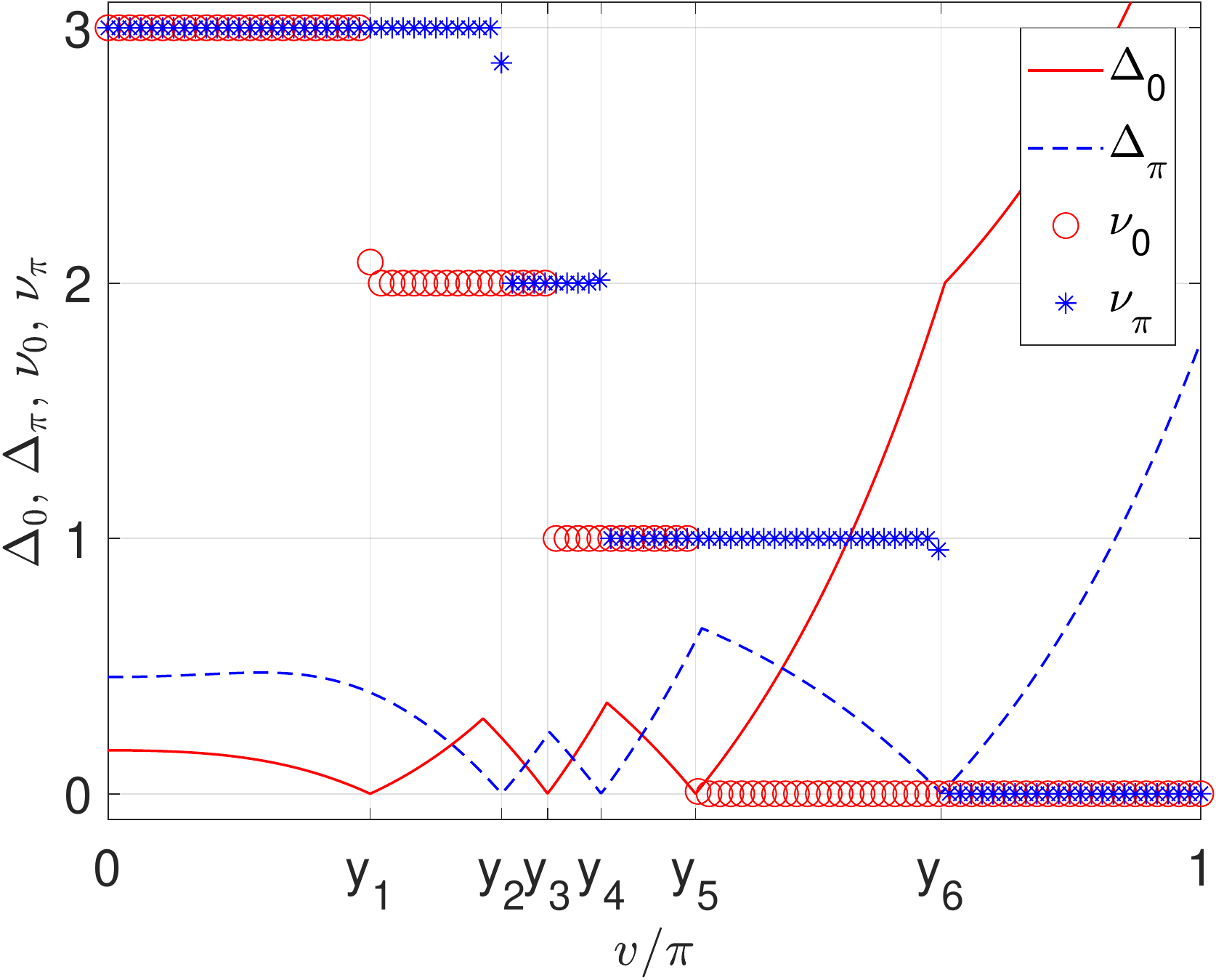}
		\par\end{centering}
	\caption{Evolution of quasienergy gap functions $\Delta_{0}$ (red solid line),
		$\Delta_{\pi}$ (blue dashed line) {[}see Eqs. (\ref{eq:GapDel0})
		and (\ref{eq:GapDelP}) for the definitions{]} and winding numbers
		$\nu_{0}$ (red circles) and $\nu_{\pi}$ (blue stars) vs. the imaginary
		parts of kicking strengths $K_{1}=u_{1}+iv$ and $K_{2}=u_{2}+iv$.
		System parameters are chosen as $u_{1}=6.5\pi$ and $u_{2}=0.5\pi$.
		The numerical values of $y_{1},y_{3},y_{5}$ ($y_{2},y_{4},y_{6}$)
		along the $v$-axis are obtained by searching the local minimum of
		the gap function $\Delta_{0}$ ($\Delta_{\pi}$) around quasienergy
		$E=0$ ($E=\pi$).\label{fig:W_vs_v}}
\end{figure}

In the first example, we choose $u_{1}=6.5\pi$ and $u_{2}=0.5\pi$
for the real parts of two kicking strengths. When $v=0$, the system
is in a Hermitian Floquet topological phase with $\nu_{0}=\nu_{\pi}=3$.
As shown in Fig. \ref{fig:W_vs_v}, increasing the imaginary kicking
strength $v$ yields consecutive Floquet topological phase transitions.
Each transition happens when one of the gap functions $(\Delta_{0},\Delta_{\pi})$ [see Eqs.~(\ref{eq:GapDel0}) and (\ref{eq:GapDelP})]
vanishes, accompanied by a quantized change of $\nu_{0}$ or $\nu_{\pi}$
by $1$. In the limit $v\rightarrow\infty$, the system becomes topologically
trivial, with $\nu_{0}=\nu_{\pi}=0$. Similar patterns of topological
phase transitions are observed by exchanging the values of $u_{1}$
and $u_{2}$ for the two kicking strengths.

\begin{figure}
	\begin{centering}
		\includegraphics[scale=0.48]{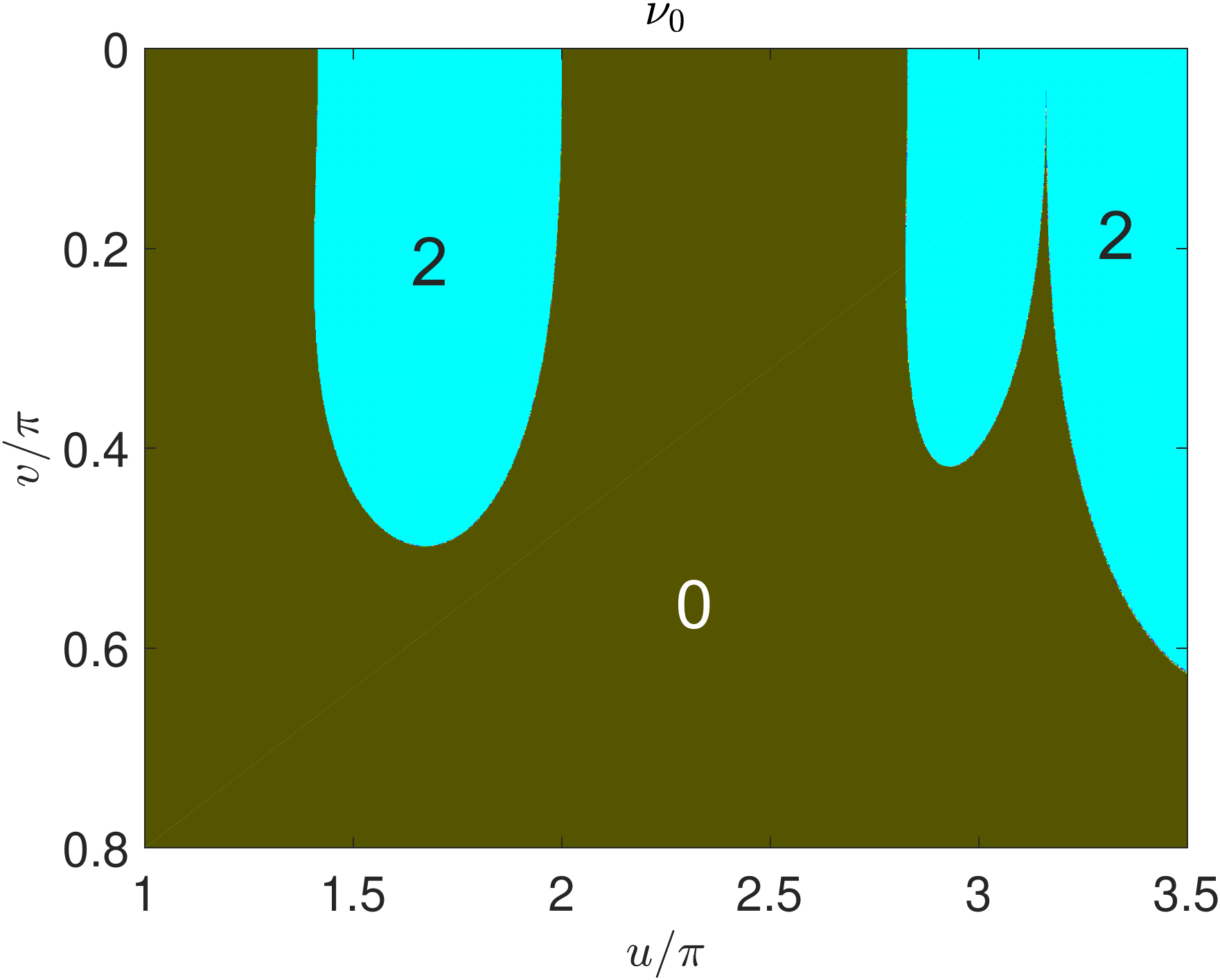}
		\par\end{centering}
	\caption{The topological phase diagram of NH-ORDKR vs. real and imaginary parts
		of kicking strengths $K_{1}=K_{2}=u+iv$. Each region with a uniform
		color corresponds to a Floquet topological phase of the NH-ORDKR,
		with the numerical value of winding number $\nu_{0}$ shown in the
		panel.\label{fig:W_vs_uv}}
\end{figure}

In the second example, we take $u_{1}=u_{2}=u$, which further indicates
that $K_{1}=K_{2}$. Plugging this condition into Eq. (\ref{eq:W0P}),
we will always have $\nu_{1}=\nu_{2}$. Therefore,
we can obtain the topological
phase diagram of the NH-ORDKR versus $u$ and $v$, with each phase
characterized only by $\nu_{0}$. A representative portion of the
phase diagram is shown in Fig. \ref{fig:W_vs_uv}. Interestingly,
we see that the increase of $u$ and $v$ could both
induce topological phase transitions in the NH-ORDKR. This further reveal the possibility
of generating rich Floquet topological states in the ORDKR by 
complex kicking potentials.

\subsubsection{Case (iii)}

\begin{figure}
	\begin{centering}
		\includegraphics[scale=0.48]{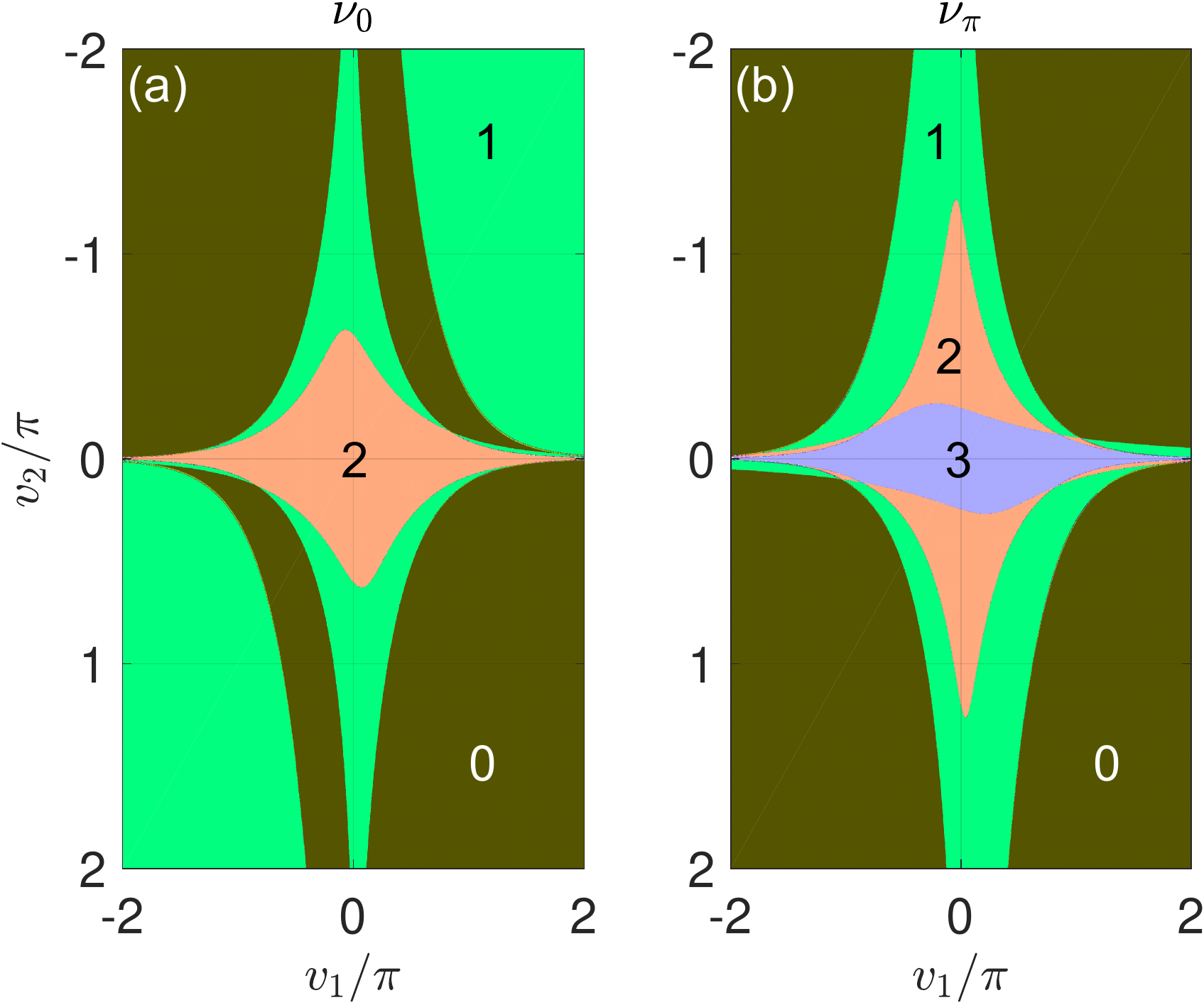}
		\par\end{centering}
	\caption{The topological phase diagram of NH-ORDKR vs. imaginary parts of kicking
		strengths $K_{1}=u_{1}+iv_{1}$ and $K_{2}=u_{2}+iv_{2}$. System
		parameters are chosen as $u_{1}=0.5\pi$ and $u_{2}=5.5\pi$. In panel
		(a) {[}(b){]}, each region with a uniform color corresponds to a Floquet
		topological phase of the NH-ORDKR, with the numerical value of winding
		number $\nu_{0}$ ($\nu_{\pi}$) shown in the panel.\label{fig:W_vs_v1v2-1}}
\end{figure}

\begin{figure}
	\begin{centering}
		\includegraphics[scale=0.48]{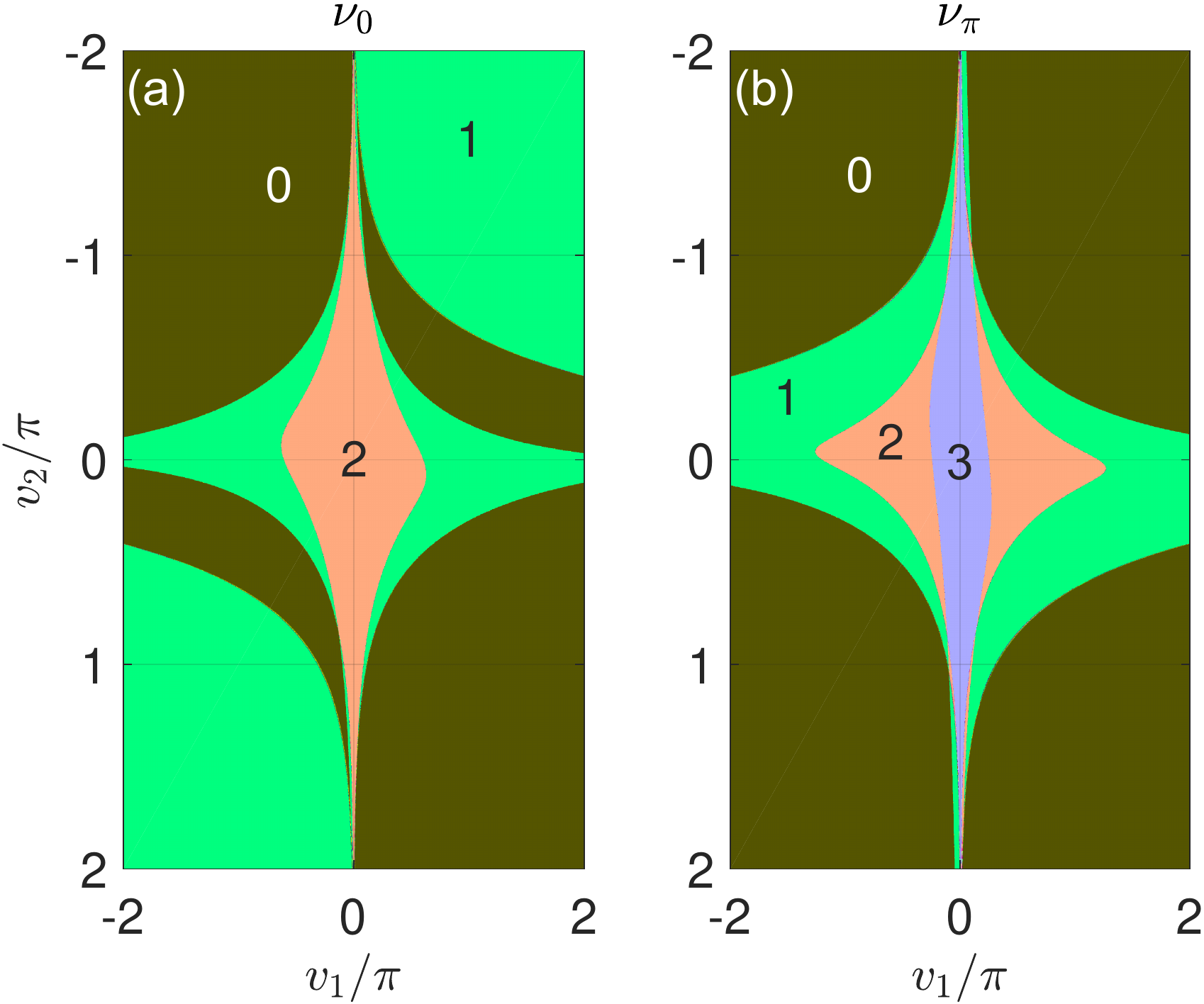}
		\par\end{centering}
	\caption{The topological phase diagram of NH-ORDKR vs. imaginary parts of kicking
		strengths $K_{1}=u_{1}+iv_{1}$ and $K_{2}=u_{2}+iv_{2}$. System
		parameters are chosen as $u_{1}=5.5\pi$ and $u_{2}=0.5\pi$. In panel
		(a) {[}(b){]}, each region with a uniform color corresponds to a Floquet
		topological phase of the NH-ORDKR, with the numerical value of winding
		number $\nu_{0}$ ($\nu_{\pi}$) shown in the panel.\label{fig:W_vs_v1v2-2}}
\end{figure}

In this case, we allow both $K_{1}$ and $K_{2}$ to be complex, with
no constraint on their imaginary parts. The resulting topological
phase diagram versus $v_{1}$ and $v_{2}$, with $(u_{1},u_{2})=(0.5\pi,5.5\pi)$
and $(u_{1},u_{2})=(5.5\pi,0.5\pi)$ are shown in Figs. \ref{fig:W_vs_v1v2-1}
and \ref{fig:W_vs_v1v2-2}, respectively. In each phase diagram, the
panels (a) and (b) correspond to the values of winding numbers $\nu_{0}$
and $\nu_{\pi}$, respectively. A region with a uniform color refers
to a parameter domain in which $\nu_{0}$ {[}panel (a){]} or $\nu_{\pi}$
{[}panel (b){]} take the same value. We see that with the change of
$v_{1}$ and $v_{2}$, a couple of non-Hermitian Floquet topological
phases are induced, with each characterized by the winding numbers
$(\nu_{0},\nu_{\pi})$. Across the boundary between two topological
phases, a quantized change of $\nu_{0}$ or $\nu_{\pi}$ is observed,
which indicates the existence of a topological phase transition.

To sum up, we find that
topological phase transitions are generic in the NH-ORDKR model, and
rich non-Hermitian Floquet topological phases could emerge under the
effect of complex kicking potentials. In the following subsection,
we will introduce a dynamical indicator -- the MCD -- to detect the winding numbers of these non-Hermitian Floquet topological phases. 

\subsection{MCD and winding numbers}\label{sec:MCD}

The MCD describes the shift of a localized wave packet in a bipartite
lattice over a long time duration. It is proposed as a way to detect the winding numbers of chiral symmetric topological insulators in one-dimension~\cite{MCD1,MCD2}. In later studies, the MCD is applied to extract the winding numbers of Floquet systems~\cite{ZhouPRA2018} and also extended to two-dimensional systems with higher order topological states~\cite{ZhouPRB2019}. In this
work, we generalize the MCD to non-Hermitian chiral symmetric Floquet systems, and using it as a dynamical probe to the winding numbers
of NH-ORDKR.

For a non-Hermitian Floquet system with chiral symmetry $\Gamma$,
we define the chiral displacement as
\begin{equation}
	C_{\alpha}(t)\equiv{\rm Tr}\left(\rho_{0}\hat{\tilde{U}}_{\alpha}^{\dagger t}\left(\hat{n}\otimes\Gamma\right)\hat{U}_{\alpha}^{t}\right),\label{eq:MCD0}
\end{equation}
where $\alpha=1,2$ is the index of symmetric time frame, $t$ is
the number of driving periods, and $\hat{n}$ is the unit cell position
operator (or momentum operator if the lattice is in momentum space).
The initial state $\rho_{0}=\frac{|0\rangle\langle0|\otimes\sigma_{0}}{2}$
describes a uniform mixture of sublattice eigenstates $|a\rangle$
and $|b\rangle$ in the $0$'s unit cell of the lattice. The choice
of $\rho_{0}$ here is different from the case in Hermitian limit,
in which the initial state occupies only a single sublattice in the
$0$'s unit cell. Furthermore, the Floquet operator $\hat{\tilde{U}}_{\alpha}$
is different from $\hat{U}_{\alpha}$ (the Floquet operator of the
system in the $\alpha$'s time frame), in the sense that if $|\psi\rangle$
is a right eigenvector of $\hat{U}_{\alpha}$ with quasienergy $E$,
then it is a left eigenvector of $\hat{\tilde{U}}_{\alpha}$ with
the same quasienergy. 

With these definitions and after relatively straightforward calculations
(see Appendix \ref{sec:App-D} for more details), the (normalized)
MCD in long-time limit is given by
\begin{alignat}{1}
	\overline{C}_{\alpha}= & \lim_{t\rightarrow\infty}\frac{1}{t}\sum_{t'=1}^{t}\int_{-\pi}^{\pi}\frac{d\theta}{2\pi}\frac{({\bf n}_{\alpha}\times\partial_{\theta}{\bf n}_{\alpha})_z}{1+|\cot(Et')|^{2}}\label{eq:MCD}\\
	= & \frac{\nu_{\alpha}}{2}.\nonumber 
\end{alignat}
Here ${\bf n}_{\alpha}=(n_{\alpha x},n_{\alpha y})$ is the winding
vector of Floquet operator in the $\alpha$'s time frame
($\alpha=1,2$). For the NH-ORDKR, explicit expressions of
$(n_{\alpha x},n_{\alpha y})$ are given by Eqs. (\ref{eq:N1X}) to
(\ref{eq:N2Y}) in Appendix \ref{sec:App-B}. 
Note that a normalization factor is introduced during the derivation of Eq.~(\ref{eq:MCD}), which helps to cancel the effects of gain/loss on the amplitude of the evolving state.
To reach the second
equality of Eq.~(\ref{eq:MCD}), we notice that $\frac{1}{1+|\cot(Et')|^{2}}=\frac{1}{2}(1-\cos[2{\rm Re}(E)t']/\cosh[2{\rm Im}(E)t'])$.
When ${\rm Im}(E)=0$, we have an oscillating factor $\frac{1}{2}(1-\cos[2{\rm Re}(E)t'])$,
which will be averaged to $\frac{1}{2}$ under $\lim_{t\rightarrow\infty}\frac{1}{t}\sum_{t'=1}^{t}$.
When ${\rm Im}(E)\neq0$, the ratio $\cos[2{\rm Re}(E)t']/\cosh[2{\rm Im}(E)t']$
will approach $0$ quickly at large $t'$, leaving only a factor $\frac{1}{2}$
in $\frac{1}{1+|\cot(Et')|^{2}}$. Therefore, we have $\lim_{t\rightarrow\infty}\frac{1}{t}\sum_{t'=1}^{t}\frac{1}{1+|\cot(Et')|^{2}}\rightarrow\frac{1}{2}$,
and the other terms under the integral of Eq. (\ref{eq:MCD}) gives
nothing but the winding number $\nu_{\alpha}$. The winding numbers
$(\nu_{0},\nu_{\pi})$ can then be obtained from $\overline{C}_{\alpha}$
as
\begin{equation}
	\nu_{0}=|\overline{C}_{1}+\overline{C}_{2}|,\qquad\nu_{\pi}=|\overline{C}_{1}-\overline{C}_{2}|.\label{eq:MCDW0P}
\end{equation}
Importantly, even though the dispersion $E(\theta)$ of NH-ORDKR is complex-valued in general, the MCD as defined in Eq.~(\ref{eq:MCD0}) could still capture the topological winding numbers of the system dynamically, which emphasize its generality as a tool in probing non-Hermitian topological phases with chiral symmetry.

\begin{figure}
	\begin{centering}
		\includegraphics[scale=0.5]{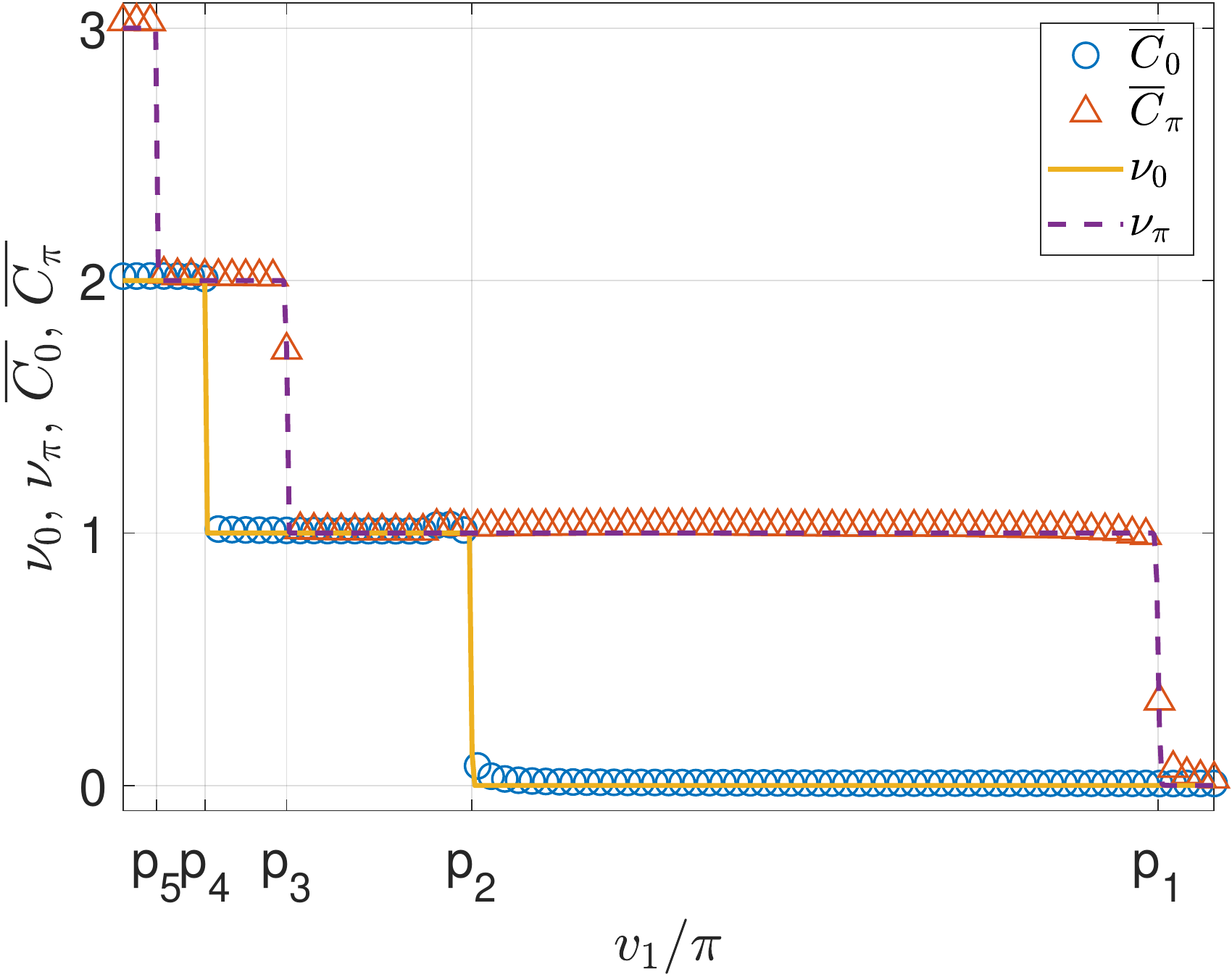}
		\par\end{centering}
	\caption{The MCD $\overline{C}_{0}=|\overline{C}_{1}+\overline{C}_{2}|$ (blue
		circles), $\overline{C}_{\pi}=|\overline{C}_{1}-\overline{C}_{2}|$
		(red triangles) and winding numbers $\nu_{0}$ (yellow solid line),
		$\nu_{\pi}$ (purple dashed line) vs. the imaginary part of kicking
		strength $K_{1}=u_{1}+iv_{1}$. System parameters are set as $u_{1}=5.5\pi$,
		$K_{2}=u_{2}=0.5\pi$, and the results for $\overline{C}_{0},\overline{C}_{\pi}$
		are averaged over $t=50$ kicking periods. $v_{1}=p_{5}$ to $p_{1}$
		correspond to gap closing points obtained from Eq. (\ref{eq:v1Cond})
		with $n=5,4,3,2,1$.\label{fig:MCD_vs_v1}}
\end{figure}

In Fig. \ref{fig:MCD_vs_v1}, we show the winding number $\nu_{0}$
(solid line) and $\nu_{\pi}$ (dashed line) of NH-ORDKR calculated
by the theoretical Eqs. (\ref{eq:W12}) and (\ref{eq:W0P}), together
with $|\overline{C}_{1}+\overline{C}_{2}|$ ($\overline{C}_{0}$ in
the figure, denoted by circles) and $|\overline{C}_{1}-\overline{C}_{2}|$
($\overline{C}_{\pi}$ in the figure, denoted by triangles) calculated
numerically by Eq. (\ref{eq:MCD}). Other system parameters are chosen
as $u_{1}=5.5\pi,u_{2}=0.5\pi$ and $v_{2}=0$. It is clear that the
theoretical predictions of $(\nu_{0},\nu_{\pi})$ and numerical results
of MCD are well consistent with each other, which verifies Eq. (\ref{eq:MCDW0P}).

\begin{figure}
	\begin{centering}
		\includegraphics[scale=0.5]{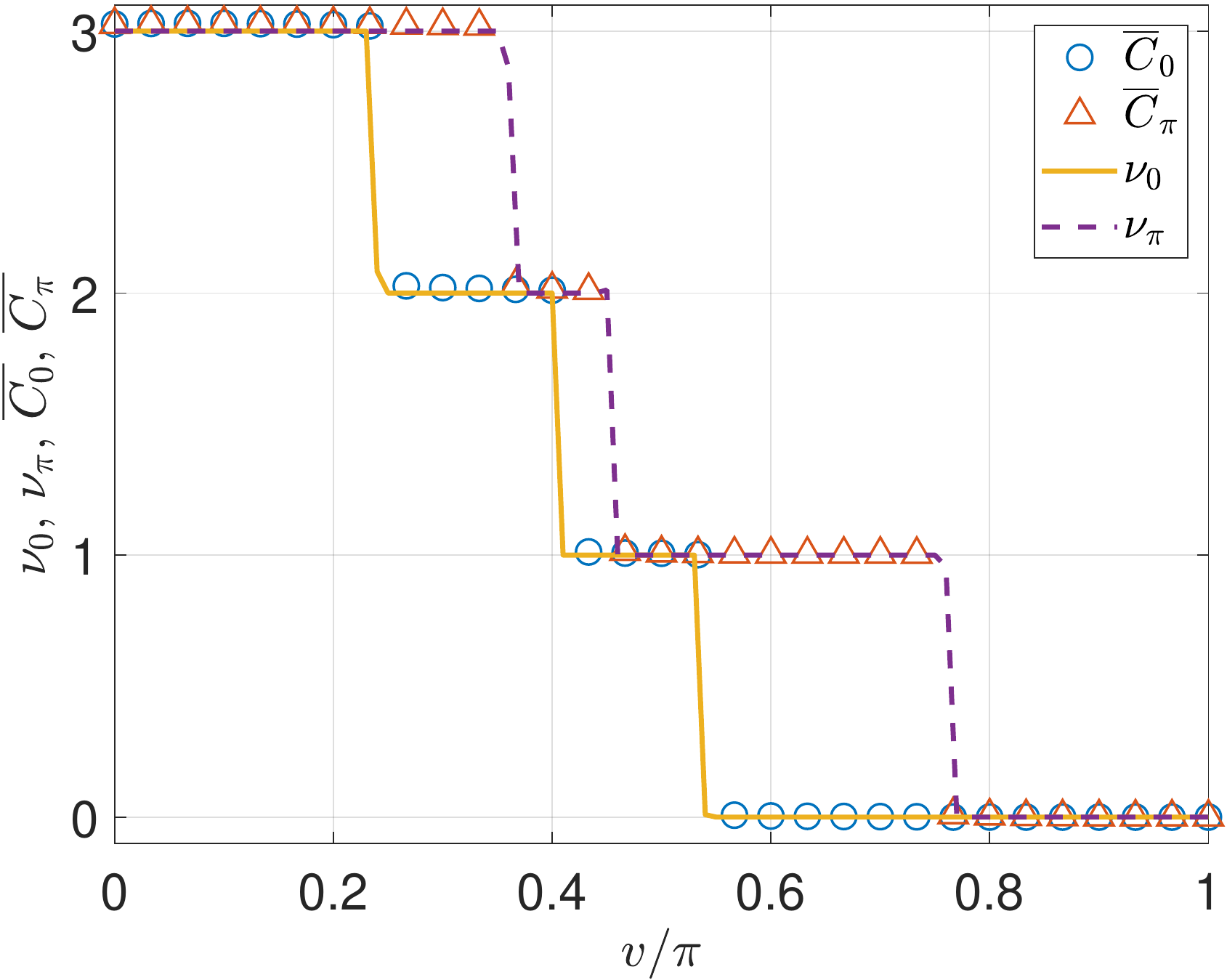}
		\par\end{centering}
	\caption{The MCD $\overline{C}_{0}=|\overline{C}_{1}+\overline{C}_{2}|$ (blue
		circles), $\overline{C}_{\pi}=|\overline{C}_{1}-\overline{C}_{2}|$
		(red triangles) and winding numbers $\nu_{0}$ (yellow solid line),
		$\nu_{\pi}$ (purple dashed line) vs. the imaginary part of kicking
		strengths $K_{1}=u_{1}+iv$ and $K_{2}=u_{2}+iv$. System parameters
		are set as $u_{1}=0.5\pi$, $u_{2}=6.5\pi$, and the results for $\overline{C}_{0},\overline{C}_{\pi}$
		are averaged over $t=50$ kicking periods.\label{fig:MCD_vs_v}}
\end{figure}

In Fig. \ref{fig:MCD_vs_v}, we give another example of MCD versus
winding numbers, in which the system parameters are $u_{1}=0.5\pi,u_{2}=6.5\pi$
and the imaginary parts of the two kicking strengths are equal. In
this case, we again observe nice consistency between the MCD and winding
numbers of NH-ORDKR within each of its topological phases. Therefore,
we conclude that the MCD, as defined by Eq. (\ref{eq:MCD}), can be
used as a generic probe to the topological winding numbers and topological
phase transitions of one-dimensional non-Hermitian Floquet systems
with chiral symmetry. To detect MCD in experiments, one may first
prepare the mixed state $\rho_{0}$ with zero momentum, and then evolve
it in two different symmetric time frames and measure the shift of
its center over different number of driving periods in each time frame.
Eqs. (\ref{eq:MCDW0P}) and (\ref{eq:MCD}) can then be used to predict
the topological winding numbers of the corresponding non-Hermitian
Floquet system.

\subsection{Edge states and bulk-edge correspondence}\label{sec:EdgeStat}
%\setcounter{equation}{0}
%\setcounter{figure}{0}
%\numberwithin{equation}{section}
%\numberwithin{figure}{section}

The bulk-edge correspondence relates the number of topological edge
states to the bulk topological invariant of the considered system.
It forms an important recipe in the characterization of topological
phases both theoretically and experimentally. The bulk-edge correspondence
in non-Hermitian systems can be more complicated~\cite{BergholtzArXiv2018,SlagerarXiv2019} due to the existence
of high-order exceptional points~\cite{HighOrderEP1,HighOrderEP2} and the so-called non-Hermitian skin effect~\cite{NHSkin2018}. In an earlier study, it has been shown that the bulk-edge correspondence
can be recovered in non-Hermitian Floquet systems under appropriate conditions~\cite{ZhouNHDQL2018}.
Below we demonstrate that the bulk-edge correspondence also hold in
the NH-ORDKR. 

For the ORDKR, the lattice is defined in momentum space, where it
is not straightforward to take an OBC and investigate
the properties of edge states. To study the bulk-edge correspondence
in the NH-ORDKR, we can map its Hamiltonian to a periodically quenched
lattice in position space. The resulting Floquet operator, according
to Eqs. (\ref{eq:UComp4}) -- (\ref{eq:UComp6}) in Appendix \ref{sec:App-A},
can be expressed as
\begin{alignat}{1}
	\hat{U}= & e^{+i\frac{\pi}{4}\sum_{n}|n\rangle\langle n|\sigma_{z}}e^{-i\frac{K_{2}}{2}\sum_{n}(|n\rangle\langle n|\sigma_{+}+|n\rangle\langle n+1|\sigma_{-}+{\rm h.c.})}\label{eq:LatticeU}\\
	\times & e^{-i\frac{\pi}{4}\sum_{n}|n\rangle\langle n|\sigma_{z}}e^{-i\frac{K_{1}}{2}\sum_{n}i(|n\rangle\langle n|\sigma_{+}+|n\rangle\langle n+1|\sigma_{-}-{\rm h.c.})},\nonumber 
\end{alignat}
where $n$ is now interpreted as the unit cell index of a real space lattice,
and the Pauli matrices operate in the space of its sublattices. In this periodically
quenched lattice model, the complex potentials $K_1$ and $K_2$ can be realized by introducing nonreciprocal hoppings and onsite gain/loss inside a unit cell and among nearest neighbor unit cells. The implementation of these effects should be within reach in current photonic-based experimental setups~\cite{XiaoNHQW2019}.

\begin{figure}
	\begin{centering}
		\includegraphics[scale=0.48]{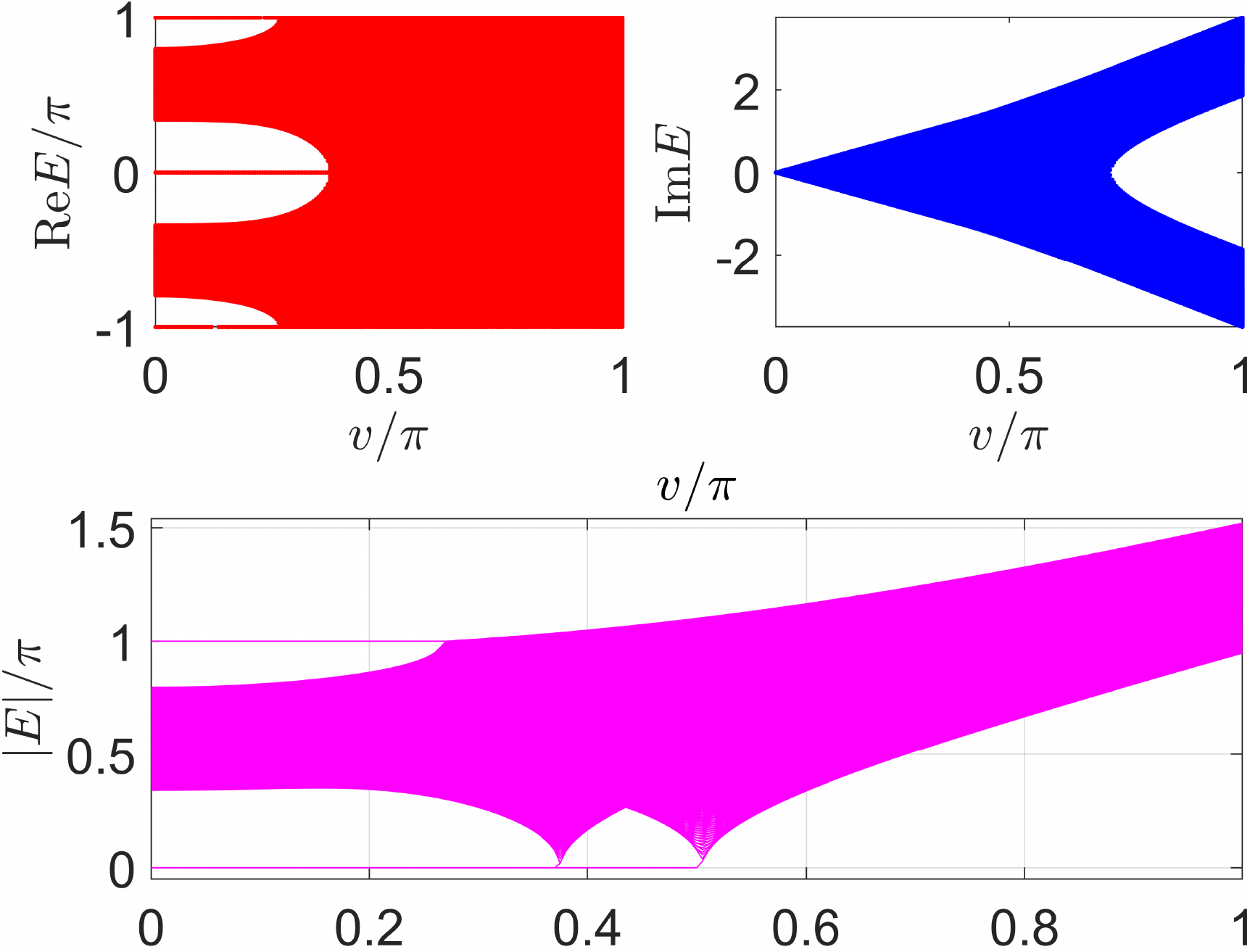}
		\par\end{centering}
	\caption{Floquet spectrum of the NH-ORDKR vs. $v$ under OBC for kicking strengths
		$K_{1}=5.5\pi+iv$ and $K_{2}=0.5\pi+iv$. The number of unit cells
		is chosen as $N=4000$ in the calculation.\label{fig:Spectrum_OBC}}
\end{figure}

The quasienergy spectrum and edge states of $\hat{U}$ can now be obtained
by solving the Floquet eigenvalue equation $\hat{U}|\psi\rangle=e^{-iE}|\psi\rangle$
under the OBC. In Fig. \ref{fig:Spectrum_OBC}, we show the Floquet spectrum of $\hat{U}$
for $u_{1}=5.5\pi,u_{2}=0.5\pi$ and $v_{1}=v_{2}=v$. In
the lower panel, we observe edge states pinned at quasienergies $0$
and $\pi$ in different regimes of the parameter space, with their
numbers change when the quasienergy gap closes at $E=0$ or $E=\pm\pi$. 

\begin{figure}
	\begin{centering}
		\includegraphics[scale=0.5]{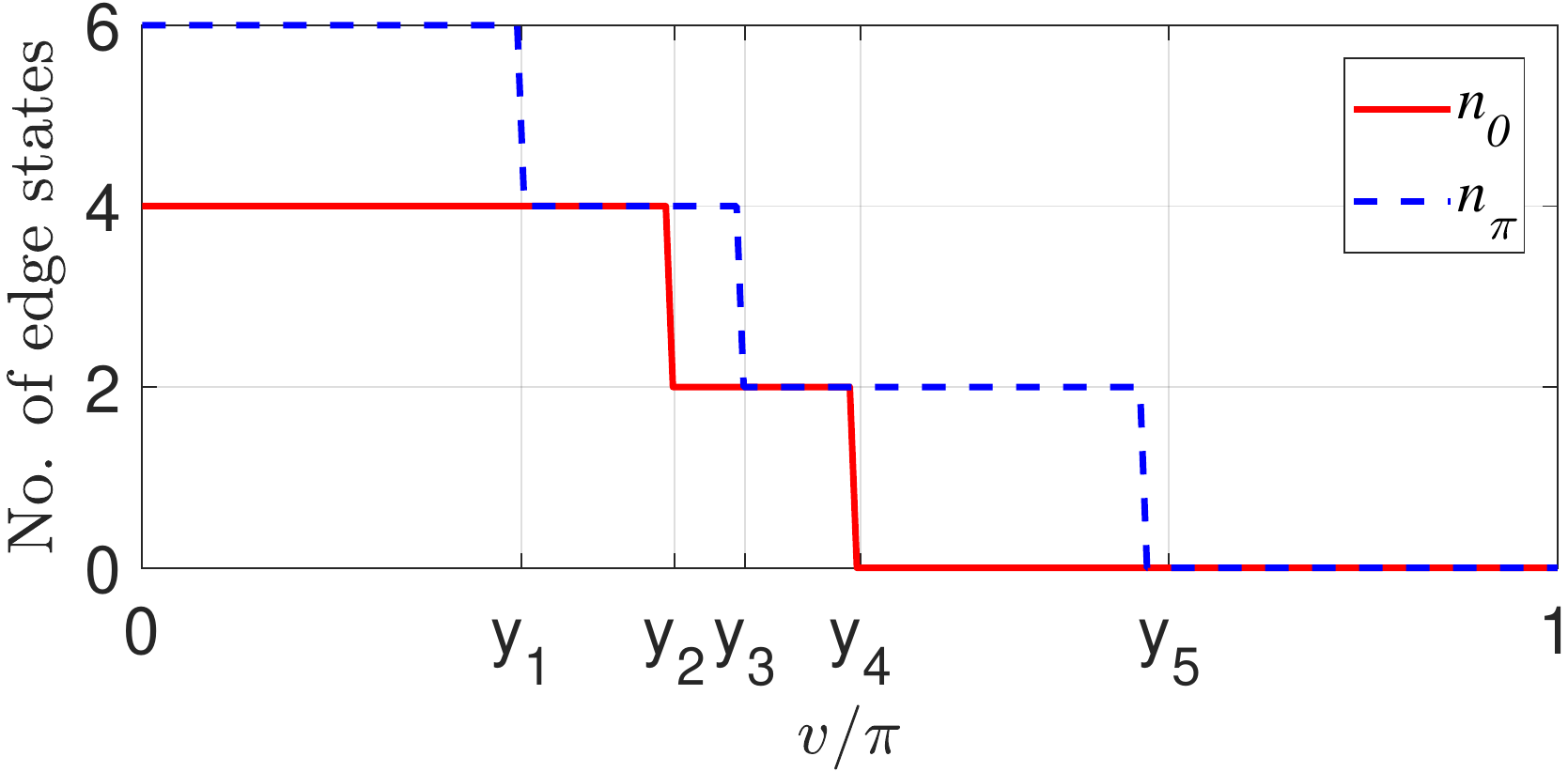}
		\par\end{centering}
	\caption{Number of edge states at quasienergies $0$ ($n_{0}$) and $\pm\pi$
		($n_{\pi}$) vs. the imaginary parts of kicking strengths $K_{1}=u_{1}+iv_{1}$
		and $K_{2}=u_{2}+iv_{2}$ in the NH-ORDKR. System parameters are chosen
		as $u_{1}=5.5\pi,u_{2}=0.5\pi$ and $v_{1}=v_{2}=v$. The number of
		unit cells is $N=4000$. The bulk Floquet spectrum is gapless at quasienergy
		$0$ or $\pi$ when $v=y_{1},y_{2},y_{3},y_{4},y_{5}$, as obtained
		from the conditions (\ref{eq:GapDel0}) and (\ref{eq:GapDelP}).\label{fig:EdgeStateNo_OBC}}
\end{figure}

In Fig. \ref{fig:EdgeStateNo_OBC}, we further show the number of
topological edge states $n_{0}$ at quasienergy $0$ (red solid line)
and $n_{\pi}$ at quasienergy $\pm\pi$ (blue dashed line) by computing
the inverse participation ratio, with the same parameter choices as in
Fig. \ref{fig:Spectrum_OBC}. $y_{1}\sim y_{5}$ along the $v$-axis
correspond to the gap closing points of the bulk quasienergy spectrum
obtained from Eq. (\ref{eq:Quasienergy}). We see that each time when
the gap closes at quasienergy $0$ ($\pi$), $n_{0}$ ($n_{\pi}$)
will get a quantized change by $2$, corresponding to a topological
phase transition with winding number $\nu_{0}$ ($\nu_{\pi}$) changing
by $1$. In other regions, the bulk-edge correspondence described
by the relations
\begin{equation}
	n_{0}=2\nu_{0},\qquad n_{\pi}=2\nu_{\pi},\label{eq:BBC}
\end{equation}
hold as in Hermitian Floquet systems, with a small deviation around
$y_{5}$ due to finite size effects. Eq. (\ref{eq:BBC}) has also
been checked numerically in other parameter regimes of the NH-ORDKR,
with similar results obtained. Therefore, we conclude that the bulk-edge
correspondence in the NH-ORDKR, as described by Eq. (\ref{eq:BBC})
holds in the same way as in the Hermitian ORDKR. Experimentally, the
non-Hermitian Floquet topological edge states have been observed in
photonic quantum walks~\cite{XiaoNHQW2019}. We expect the relation (\ref{eq:BBC}) of
NH-ORDKR to be verifiable in similar experimental setups.

\section{Conclusions}\label{summary}

In this work, we investigated Floquet topological phases in a non-Hermitian
extension of the double kicked rotor, which is
a prototypical example of dynamical kicking systems. Under the on
resonance condition, the system possesses rich non-Hermitian
Floquet topological phases protected by chiral symmetry. The topological phase diagram 
of the NH-ORDKR is obtained, with each of its phase
being characterized by a pair of integer winding numbers. These winding
numbers could be detected dynamically by measuring the mean
chiral displacement in two symmetric time frames. Furthermore, by mapping
our model to a periodically quenched lattice, we found its topological
edge states. The number of these states at quasienergies $0$ and
$\pm\pi$ in each topological phase is precisely counted by the winding
numbers of bulk states, revealing the bulk-edge correspondence of the NH-ORDKR. 

In future studies, more fruitful topological structures
are expected to appear in non-Hermitian dynamical kicking systems after introducing
spin degrees of freedom and many-body interactions.
Interesting examples include the recently discovered Floquet non-Hermitian skin effect~\cite{NHSkin12} in momentum space and the non-Hermitian counterpart of Floquet topological time crystals~\cite{BomantaraFloquetQC2018}.
New schemes that
go beyond the existing $38$-fold way for classification of static non-Hermitian topological
phases~\cite{NHTIClassification1,NHTIClassification2} should be required to achieve a full characterization of these non-Hermitian Floquet states~\cite{ZhouarXiv2019}. 
On application side, the mean chiral displacement proposed in this work could assist the future experimental detection of topological invariants in non-Hermitian Floquet systems.
Furthermore, with the promising proposal of Floquet topological quantum computing~\cite{BomantaraFloquetQC2018} and Floquet quantum state transfer~\cite{TanarXiv2019}, it would be interesting to investigate the potential of the Floquet topological edge states found in this work in achieving quantum computing and quantum information transfer against environmental effects that can be modeled by non-Hermitian Hamiltonians.

\section*{Acknowledgement}
L. Z. acknowledges Jiangbin Gong and Bo Qu for helpful comments. This work is supported by the National Natural Science Foundation of China (Grant No.~11905211), the Young Talents Project at Ocean University of China (Grants No.~861801013196 and 841912009) and the Applied Research Project of Postdoctoral Fellows in Qingdao (Grant No.~861905040009).

\appendix
%\vspace{0.5cm}

\section{Floquet operator in different representations}\label{sec:App-A}
The Floquet operator of NH-ORDKR, as given by Eq. (\ref{eq:NH-ORDKR})
in the main text, can be expressed in the momentum lattice representation
as follows. We first write out its component terms as
\begin{alignat}{1}
	& e^{\pm\frac{\pi}{2}\hat{n}^{2}}\label{eq:UComp1}\\
	= & e^{\pm i\frac{\pi}{4}}e^{\pm i\frac{\pi}{4}\sum_{\ell}(|2\ell-1\rangle\langle2\ell-1|-|2\ell\rangle\langle2\ell|)},\nonumber \\
	& e^{-iK_{1}\cos(\hat{x}+\beta)}\label{eq:UComp2}\\
	= & e^{-i\frac{K_{1}}{2}\sum_{\ell}(e^{i\beta}|2\ell-1\rangle\langle2\ell|+e^{i\beta}|2\ell\rangle\langle2\ell+1|+{\rm h.c.})},\nonumber \\
	& e^{-iK_{2}\cos(\hat{x})}\label{eq:UComp3}\\
	= & e^{-i\frac{K_{2}}{2}\sum_{\ell}(|2\ell-1\rangle\langle2\ell|+|2\ell\rangle\langle2\ell+1|+{\rm h.c.})},\nonumber 
\end{alignat}
where the resolution identity $I=\sum_{\ell}|\ell\rangle\langle\ell|$
has been inserted to arrive at the expansions. Since Eq. (\ref{eq:NH-ORDKR})
is invariant under the translation over two sites in momentum space,
we could decompose the momentum space lattice into two chains containing
only odd and even sites, denoted by sublattice indices $a$ and $b$,
respectively. A unit cell of the momentum space lattice now contains
two sublattice sites, and we can introduce Pauli matrices in the sublattice
representation as
\begin{alignat}{1}
	\sigma_{x}= & |a\rangle\langle b|+|b\rangle\langle a|,\\
	\sigma_{y}= & i(|b\rangle\langle a|-|a\rangle\langle b|),\\
	\sigma_{z}= & |a\rangle\langle a|-|b\rangle\langle b|.
\end{alignat}
The sublattice raising and lower operators can also be expressed as
\begin{equation}
	\sigma_{\pm}=\frac{\sigma_{x}+i\sigma_{y}}{2}.\label{eq:SRSL}
\end{equation}
In this bipartite lattice representation, Eqs. (\ref{eq:UComp1})
to (\ref{eq:UComp3}) can be written as
\begin{alignat}{1}
	& e^{\pm\frac{\pi}{2}\hat{n}^{2}}=e^{\pm i\frac{\pi}{4}}e^{\pm i\frac{\pi}{4}\sum_{n}|n\rangle\langle n|\sigma_{z}}\label{eq:UComp4}\\
	& e^{-iK_{1}\cos(\hat{x}+\beta)}\label{eq:UComp5}\\
	= & e^{-i\frac{K_{1}}{2}\sum_{n}(e^{i\beta}|n\rangle\langle n|\sigma_{+}+e^{i\beta}|n\rangle\langle n+1|\sigma_{-}+{\rm h.c.})},\nonumber \\
	& e^{-iK_{2}\cos(\hat{x})}\label{eq:UComp6}\\
	= & e^{-i\frac{K_{2}}{2}\sum_{n}(|n\rangle\langle n|\sigma_{+}+|n\rangle\langle n+1|\sigma_{-}+{\rm h.c.})},\nonumber 
\end{alignat}
where $n$ is the unit cell index. Performing the Fourier transforms
$|n\rangle=\sum_{\theta}e^{-in\theta}|\theta\rangle$, $\langle n|=\sum_{\theta}e^{in\theta}\langle\theta|$
to Eqs. (\ref{eq:UComp4}) to (\ref{eq:UComp6}) and choosing $\beta=\frac{\pi}{2}$,
we arrive at
\begin{alignat}{1}
	& e^{\pm\frac{\pi}{2}\hat{n}^{2}}=e^{\pm i\frac{\pi}{4}}e^{\pm i\frac{\pi}{4}\sum_{\theta}|\theta\rangle\langle\theta|\sigma_{z}}\\
	& e^{-iK_{1}\cos(\hat{x}+\beta)}\\
	= & e^{-i\frac{K_{1}}{2}\sum_{\theta}|\theta\rangle\langle\theta|(i\sigma_{+}+ie^{i\theta}\sigma_{-}+{\rm h.c.})},\nonumber \\
	& e^{-iK_{2}\cos(\hat{x})}\\
	= & e^{-i\frac{K_{2}}{2}\sum_{\theta}|\theta\rangle\langle\theta|(\sigma_{+}+e^{i\theta}\sigma_{-}+{\rm h.c.})}.\nonumber 
\end{alignat}
Combing these terms in sequential order and using Eq. (\ref{eq:SRSL}),
we obtain the Floquet operator of NH-ORDKR in the form $\hat{U}=\sum_{\theta}U(\theta)|\theta\rangle\langle\theta|$,
with
\begin{alignat}{1}
	U(\theta)= & e^{+i\frac{\pi}{4}\sigma_{z}}e^{-i\frac{K_{2}}{2}[(1+\cos\theta)\sigma_{x}+\sin\theta\sigma_{y}]}\nonumber \\
	\times & e^{-i\frac{\pi}{4}\sigma_{z}}e^{+i\frac{K_{1}}{2}[\sin\theta\sigma_{x}+(1-\cos\theta)\sigma_{y}]}\label{eq:UTheta0}
\end{alignat}
Finally, using trigonometric relations $\sin\theta=2\sin\frac{\theta}{2}\cos\frac{\theta}{2}$
and $\cos\theta=2\cos^{2}\frac{\theta}{2}-1=1-2\sin^{2}\frac{\theta}{2}$,
we arrive at Eq. (\ref{eq:UTheta}) of the main text.

$U(\theta)$ can be further expressed in the two symmetric time frames
as discussed in the main text. To do so, we first shift the starting
time of the evolution to the start of the second half of driving period,
and split the kick $e^{-i\frac{K_{2}}{2}[(1+\cos\theta)\sigma_{x}+\sin\theta\sigma_{y}]}$
into two `` half kicks'' at the start and end of the shifted evolution.
The resulting Floquet operator in this new time frame is given by
\begin{alignat}{1}
	U_{1}(\theta)= & e^{-i\frac{{\cal K}_{2}}{2}\left(\cos\frac{\theta}{2}\sigma_{x}+\sin\frac{\theta}{2}\sigma_{y}\right)}\nonumber \\
	\times & e^{-i\frac{\pi}{4}\sigma_{z}}e^{+i{\cal K}_{1}\left(\cos\frac{\theta}{2}\sigma_{x}+\sin\frac{\theta}{2}\sigma_{y}\right)}e^{+i\frac{\pi}{4}\sigma_{z}}\label{eq:U1Theta0}\\
	\times & e^{-i\frac{{\cal K}_{2}}{2}\left(\cos\frac{\theta}{2}\sigma_{x}+\sin\frac{\theta}{2}\sigma_{y}\right)}.\nonumber 
\end{alignat}
Similarly, by splitting $e^{+i{\cal K}_{1}\left(\cos\frac{\theta}{2}\sigma_{x}+\sin\frac{\theta}{2}\sigma_{y}\right)}$
into two ``half kicks'' and shifting one of them to the end of the
evolution, $U(\theta)$ in Eq. (\ref{eq:UTheta0}) becomes
\begin{alignat}{1}
	U_{2}(\theta)= & e^{+i\frac{{\cal K}_{1}}{2}\left(\cos\frac{\theta}{2}\sigma_{x}+\sin\frac{\theta}{2}\sigma_{y}\right)}\nonumber \\
	\times & e^{+i\frac{\pi}{4}\sigma_{z}}e^{-i{\cal K}_{2}\left(\cos\frac{\theta}{2}\sigma_{x}+\sin\frac{\theta}{2}\sigma_{y}\right)}e^{-i\frac{\pi}{4}\sigma_{z}}\label{eq:U2Theta0}\\
	\times & e^{+i\frac{{\cal K}_{1}}{2}\left(\cos\frac{\theta}{2}\sigma_{x}+\sin\frac{\theta}{2}\sigma_{y}\right)}.\nonumber 
\end{alignat}
It is clear that both $U_{1}(\theta)$ and $U_{2}(\theta)$ are related
to $U(\theta)$ by similarity transformations. Finally, using the
transformations $e^{\mp i\frac{\pi}{4}\sigma_{z}}\sigma_{x}e^{\pm i\frac{\pi}{4}\sigma_{z}}=\pm\sigma_{y}$
and $e^{\mp i\frac{\pi}{4}\sigma_{z}}\sigma_{y}e^{\pm i\frac{\pi}{4}\sigma_{z}}=\mp\sigma_{x}$,
Eqs. (\ref{eq:U1Theta0}) and (\ref{eq:U2Theta0}) simplify to Eqs.
(\ref{eq:U1Theta}) and (\ref{eq:U2Theta}) in the main text.

\section{Explicit expressions of the Floquet operators}\label{sec:App-B}

Using the Euler formula $e^{i\phi{\bf n}\cdot\boldsymbol{\sigma}}=\cos\phi+i\sin\phi{\bf n}\cdot\boldsymbol{\sigma}$,
we can expand each exponential of Eqs. (\ref{eq:U1Theta}) and (\ref{eq:U2Theta})
in the main text. The resulting terms can be recombined to give
\begin{alignat}{1}
	U_{1}(\theta)= & \cos{\cal K}_{1}\cos{\cal K}_{2}\label{eq:U1ThetaEP}\\
	-i & \left[\cos\frac{\theta}{2}\cos{\cal K}_{1}\sin{\cal K}_{2}+\sin\frac{\theta}{2}\sin{\cal K}_{1}\right]\sigma_{x}\nonumber \\
	-i & \left[\sin\frac{\theta}{2}\cos{\cal K}_{1}\sin{\cal K}_{2}-\cos\frac{\theta}{2}\sin{\cal K}_{1}\right]\sigma_{y},\nonumber 
\end{alignat}
and
\begin{alignat}{1}
	U_{2}(\theta)= & \cos{\cal K}_{1}\cos{\cal K}_{2}\label{eq:U2ThetaEP}\\
	-i & \left[-\cos\frac{\theta}{2}\sin{\cal K}_{1}\cos{\cal K}_{2}+\sin\frac{\theta}{2}\sin{\cal K}_{2}\right]\sigma_{x}\nonumber \\
	-i & \left[-\sin\frac{\theta}{2}\sin{\cal K}_{1}\cos{\cal K}_{2}-\cos\frac{\theta}{2}\sin{\cal K}_{2}\right]\sigma_{y},\nonumber 
\end{alignat}
By setting $\cos[E(\theta)]=\cos{\cal K}_{1}\cos{\cal K}_{2}$, it
is straightforward to see that $E(\theta)=\arccos(\cos{\cal K}_{1}\cos{\cal K}_{2})$,
and Eqs. (\ref{eq:U1ThetaEP}) and (\ref{eq:U2ThetaEP}) has the form
of Eq. (\ref{eq:UAlphaTheta}) in the main text, with
\begin{alignat}{1}
	n_{1x}= & \frac{+\cos\frac{\theta}{2}\cos{\cal K}_{1}\sin{\cal K}_{2}+\sin\frac{\theta}{2}\sin{\cal K}_{1}}{\sin E(\theta)},\label{eq:N1X}\\
	n_{1y}= & \frac{+\sin\frac{\theta}{2}\cos{\cal K}_{1}\sin{\cal K}_{2}-\cos\frac{\theta}{2}\sin{\cal K}_{1}}{\sin E(\theta)},\label{eq:N1Y}\\
	n_{2x}= & \frac{-\cos\frac{\theta}{2}\sin{\cal K}_{1}\cos{\cal K}_{2}+\sin\frac{\theta}{2}\sin{\cal K}_{2}}{\sin E(\theta)},\label{eq:N2X}\\
	n_{2y}= & \frac{-\sin\frac{\theta}{2}\sin{\cal K}_{1}\cos{\cal K}_{2}-\cos\frac{\theta}{2}\sin{\cal K}_{2}}{\sin E(\theta)}.\label{eq:N2Y}
\end{alignat}

\section{Gapless conditions}\label{sec:App-C}

We present derivation details for the gap closing conditions. Using
shorthand notations
\begin{alignat}{1}
	\mathfrak{u}_{1}\equiv u_{1}\sin\frac{\theta}{2}, & \qquad\mathfrak{u}_{2}\equiv u_{2}\cos\frac{\theta}{2},\\
	\mathfrak{v}_{1}\equiv v_{1}\sin\frac{\theta}{2}, & \qquad\mathfrak{v}_{2}\equiv v_{2}\cos\frac{\theta}{2},
\end{alignat}
we can express the gap closing condition as
\begin{equation}
	\cos E=\cos(\mathfrak{u}_{1}+i\mathfrak{v}_{1})\cos(\mathfrak{u}_{2}+i\mathfrak{v}_{2})=\pm1.\label{eq:GapCond0}
\end{equation}
When $v_{1}\neq0$ and $v_{2}=0$, this condition is equivalent to:
\begin{alignat}{1}
	\cos\mathfrak{u}_{1}\cosh\mathfrak{v}_{1}\cos\mathfrak{u}_{2}= & \pm1,\label{eq:GPCRe1}\\
	\sin\mathfrak{u}_{1}\sinh\mathfrak{v}_{1}\cos\mathfrak{u}_{2}= & 0.\label{eq:GPCIm1}
\end{alignat}
It is clear that to satisfy both of the equations, $\cos\mathfrak{u}_{2}$
cannot be zero. Furthermore, if $\sinh\mathfrak{v}_{1}=0$, we must
have $\sin\frac{\theta}{2}=0$, and Eq. (\ref{eq:GPCIm1}) will be
satisfied only if $\cos(u_{2})=\pm1$, which is a very special condition
that is irrelevant to the value of $v_{1}$. Therefore, Eq. (\ref{eq:GPCIm1})
is generally satisfied if $\sin\mathfrak{u}_{1}=0$, yielding $\sin\frac{\theta}{2}=\frac{n\pi}{u_{1}}$
for $n\pi\leq u_{1}$ with $n\in\mathbb{N}$. Plugging this relation
into Eq. (\ref{eq:GPCRe1}) and regroup the relevant terms, we obtain
Eq. (\ref{eq:v1Cond}) in the main text. Eq. (\ref{eq:v2Cond}) can
be derived in a similar manner.

In more general situations, the gapless condition can be extracted
numerically. We first separate Eq. (\ref{eq:GapCond0}) into its real
part $f$ and imaginary part $g$. Expressed in terms of $f$ and
$g$, the Floquet spectrum is gapless when
\begin{alignat}{1}
	\pm1=f= & \cos\mathfrak{u}_{1}\cos\mathfrak{u}_{2}\cosh\mathfrak{v}_{1}\cosh\mathfrak{v}_{2}\nonumber \\
	- & \sin\mathfrak{u}_{1}\sin\mathfrak{u}_{2}\sinh\mathfrak{v}_{1}\sinh\mathfrak{v}_{2},
\end{alignat}
and
\begin{alignat}{1}
	0=g= & \cos\mathfrak{u}_{1}\sin\mathfrak{u}_{2}\cosh\mathfrak{v}_{1}\sinh\mathfrak{v}_{2}\nonumber \\
	+ & \sin\mathfrak{u}_{1}\cos\mathfrak{u}_{2}\sinh\mathfrak{v}_{1}\cosh\mathfrak{v}_{2}.
\end{alignat}
Using $f$ and $g$, we could further introduce a pair of functions
$(\Delta_{0},\Delta_{\pi})$ to characterize the size of spectrum
gaps at quasienergy $E=0$ and $E=\pm\pi$, respectively. Explicitly,
these functions are defined as
\begin{alignat}{1}
	\Delta_{0}= & \sqrt{(f-1)^{2}+g^{2}},\label{eq:GapDel0}\\
	\Delta_{\pi}= & \sqrt{(f+1)^{2}+g^{2}}.\label{eq:GapDelP}
\end{alignat}
Therefore, the spectrum becomes gapless at the center (edge) of the
quasienergy Brillouin zone if $\Delta_{0}=0$ ($\Delta_{\pi}=0$). 

\section{Derivation of the mean chiral displacement}\label{sec:App-D}

We provide derivation details for Eq. (\ref{eq:MCD}) in this appendix.
In the definition of chiral displacement by Eq. (\ref{eq:MCD0}),
we can insert the identity in lattice representation to yield
\begin{alignat}{1}
	& {\rm Tr}\left(\rho_{0}\hat{\tilde{U}}_{\alpha}^{\dagger t}\left(\hat{n}\otimes\Gamma\right)\hat{U}_{\alpha}^{t}\right)=\frac{1}{2}\sum_{n}\sum_{s,s'=a,b}\nonumber \\
	\times & n\langle0|\langle s|\hat{\tilde{U}}_{\alpha}^{\dagger t}|n\rangle\Gamma|s'\rangle\langle s'|\langle n|\hat{U}_{\alpha}^{t}|0\rangle|s\rangle.
\end{alignat}
Expressing $\hat{U}_{\alpha}$ and $\hat{\tilde{U}}_{\alpha}^{\dagger}$
in quasiposition (or quasimomentum for real space lattices) representation
as $\hat{U}_{\alpha}=\sum_{\theta}|\theta\rangle U_{\alpha}(\theta)\langle\theta|$
and $\hat{\tilde{U}}_{\alpha}^{\dagger}=\sum_{\theta}|\theta\rangle\tilde{U}_{\alpha}^{\dagger}(\theta)\langle\theta|$,
with $U_{\alpha}(\theta)$ and $\tilde{U}_{\alpha}^{\dagger}(\theta)$
being $2\times2$ matrices in the sublattice representation, we further
obtain
\begin{alignat}{1}
	& {\rm Tr}\left(\rho_{0}\hat{\tilde{U}}_{\alpha}^{\dagger t}\left(\hat{n}\otimes\Gamma\right)\hat{U}_{\alpha}^{t}\right)=\frac{1}{2}\sum_{n}\sum_{\theta\theta'}n\nonumber \\
	\times & \langle0|\theta\rangle\langle\theta|n\rangle\langle n|\theta'\rangle\langle\theta'|0\rangle{\rm Tr}\left[\tilde{U}_{\alpha}^{\dagger t}(\theta)\Gamma U_{\alpha}^{t}(\theta')\right],
\end{alignat}
where the trace is now taken over the sublattice degrees of freedom.
Using the Fourier transform relations
\begin{alignat}{1}
	|\theta\rangle= & \frac{1}{\sqrt{N}}\sum_{n}e^{i\theta n}|n\rangle,\nonumber \\
	|n\rangle= & \frac{1}{\sqrt{N}}\sum_{n}e^{-i\theta n}|n\rangle,\\
	\langle n|\theta\rangle= & \frac{1}{\sqrt{N}}e^{i\theta n},\nonumber 
\end{alignat}
we can simplify the numerator to
\begin{alignat}{1}
	& {\rm Tr}\left(\rho_{0}\hat{\tilde{U}}_{\alpha}^{\dagger t}\left(\hat{n}\otimes\Gamma\right)\hat{U}_{\alpha}^{t}\right)=\frac{1}{2}\sum_{n}\sum_{\theta\theta'}n\nonumber \\
	\times & \frac{1}{N^{2}}e^{in(\theta'-\theta)}{\rm Tr}\left[\tilde{U}_{\alpha}^{\dagger t}(\theta)\Gamma U_{\alpha}^{t}(\theta')\right],
\end{alignat}
Using the relation
\begin{equation}
	\frac{1}{N}\sum_{n}ne^{i(\theta'-\theta)n}=i\partial_{\theta}\frac{1}{N}\sum_{n}e^{i(\theta'-\theta)n}=i\partial_{\theta}\delta_{\theta\theta'},
\end{equation}
we find
\begin{alignat}{1}
	& {\rm Tr}\left(\rho_{0}\hat{\tilde{U}}_{\alpha}^{\dagger t}\left(\hat{n}\otimes\Gamma\right)\hat{U}_{\alpha}^{t}\right)=\frac{1}{2}\sum_{\theta\theta'}\nonumber \\
	\times & \frac{1}{N}i\partial_{\theta}\delta_{\theta\theta'}{\rm Tr}\left[\tilde{U}_{\alpha}^{\dagger t}(\theta)\Gamma U_{\alpha}^{t}(\theta')\right],
\end{alignat}
In the continuous limit ($N\rightarrow\infty$), we have $\delta_{\theta\theta'}\rightarrow\frac{2\pi}{N}\delta(\theta-\theta')$
and $\sum_{\theta,\theta'}\rightarrow N^{2}\int_{-\pi}^{\pi}\frac{d\theta}{2\pi}\int_{-\pi}^{\pi}\frac{d\theta'}{2\pi}$.
Combining this into the Eq.~(D6) above then leads to
\begin{alignat}{1}
	C_{\alpha}(t)= & {\rm Tr}\left(\rho_{0}\hat{\tilde{U}}_{\alpha}^{\dagger t}\left(\hat{n}\otimes\Gamma\right)\hat{U}_{\alpha}^{t}\right)\\
	= & \frac{1}{2}\int_{-\pi}^{\pi}\frac{d\theta}{2\pi}{\rm Tr}\left[\tilde{U}_{\alpha}^{\dagger t}(\theta)\Gamma i\partial_{\theta}U_{\alpha}^{t}(\theta)\right].\nonumber 
\end{alignat}
Inserting the normalization factor $\frac{1}{2}{\rm Tr}\left[\tilde{U}_{\alpha}^{\dagger t}(\theta)U_{\alpha}^{t}(\theta)\right]$
at each $\theta$ (since the evolution will change the normal of the
state), and taking the long time average $\lim_{t\rightarrow\infty}\frac{1}{t}\sum_{t'=1}^{t}$,
we obtain the expression for MCD as
\begin{equation}
	\overline{C}_{\alpha}=\lim_{t\rightarrow\infty}\frac{1}{t}\sum_{t'=1}^{t}\int_{-\pi}^{\pi}\frac{d\theta}{2\pi}\frac{{\rm Tr}\left[\tilde{U}_{\alpha}^{\dagger t'}(\theta)\Gamma i\partial_{\theta}U_{\alpha}^{t'}(\theta)\right]}{{\rm Tr}\left[\tilde{U}_{\alpha}^{\dagger t'}(\theta)U_{\alpha}^{t'}(\theta)\right]}\label{eq:MCDApp}
\end{equation}

For the NH-ORDKR, we have $\Gamma=\sigma_{z}$, $U_{\alpha}(\theta)=e^{-iE({\bf n}_{\alpha}\cdot\boldsymbol{\sigma})}$
and $\tilde{U}_{\alpha}^{\dagger}(\theta)=e^{+iE^{*}({\bf n}_{\alpha}\cdot\boldsymbol{\sigma})}$.
Plugging these into Eq. (\ref{eq:MCDApp}), the numerator and denominator
become
\begin{alignat}{1}
	{\rm Tr}\left[\tilde{U}_{\alpha}^{\dagger t'}(\theta)U_{\alpha}^{t'}(\theta)\right]= & 2\left[|\cos(Et')|^{2}+|\sin(Et')|^{2}\right]\\
	{\rm Tr}\left[\tilde{U}_{\alpha}^{\dagger t'}(\theta)\Gamma i\partial_{\theta}U_{\alpha}^{t'}(\theta)\right]= & 2|\sin(Et')|^{2}\left({\bf n}_{\alpha}\times\partial_{\theta}{\bf n}_{\alpha}\right)_{z}
\end{alignat}
Combining these into Eq. (\ref{eq:MCDApp}), we finally obtain Eq.
(\ref{eq:MCD}) in the main text.

%\color{black}
%\newpage{}

\end{document}